\documentclass[11pt,a4paper]{article}
\usepackage{amssymb}

\usepackage{graphicx}
\usepackage{amsmath}
\usepackage{mathpazo}
\usepackage{euscript}
\usepackage[pdftex]{hyperref}

\numberwithin{equation}{section}
\makeatletter\@addtoreset{equation}{section}

\title{\vspace{-2mm}\selectfont \textbf{New coherent states with Laguerre polynomials coefficients for the symmetric P\"{o}schl-Teller oscillator}}
\author{\textsc{PATRICK KAYUPE KIKODIO$^{\flat}$ and ZOUHA\"{I}R MOUAYN$^{\natural}$}\\
\small ${}^{\flat}$13, Market avenue, Mongt-Ngafula, Kinshasa, Congo (DRC)\\
\small ${}^{\natural}$Department of Mathematics, Faculty of Sciences and Technics (M'Ghila)\\ \small P.O.Box. 523, B\'{e}ni Mellal, Morocco}
\date{}
\begin{document}
\maketitle
\begin{abstract}
We construct a new class of coherent states labeled by points $z$ of the complex plane and depending on three numbers $(\gamma,\nu)$ and $\varepsilon>0$ by replacing the coefficients $z^{n}/\sqrt{n!}$ of the canonical coherent states by Laguerre polynomials.
These states are superpositions of eigenstates of the symmetric P\"{o}schl-Teller
oscillator and they solve the identity of the states Hilbert space at the limit $\varepsilon \rightarrow 0^{+}$. Their wavefunctions are obtained in a closed form for a special case of parameters $(\gamma, \nu)$. We discuss
 their associated coherent states transform which leads to an integral representation of Hankel type for Laguerre functions.
\end{abstract}
\section{Introduction}

The recent works on quantum dots wells in nanophysics [1-2] strongly motivate
the construction of quantum states for infinite wells with localization properties
comparable to those of Schr\"{o}dinger states. Infinite wells are often modeled by a
family of P\"{o}schl-Teller (PT) potentials [3]. The symmetric P\"{o}schl-Teller (SPT)
potentials are a subclass of PT potentials, which are widely used in molecular
and solid state physics. Many other potentials can be obtained from PT potentials
by appropriate limiting procedure and point canonical transformations.

There is an extensive literature in which many characteristic properties of the PT potential
were examined at classical mechanics as well as at quantum mechanics levels.
Another interesting aspect of this potential lies in the fact that it has a quadratic
spectrum leading to a rich revival structure for its coherent states, which makes
possible the formation of Schr\"{o}dinger cat and cat-like states. Different types of
coherent states for quantum mechanical systems evolving in the PT potential
have been discussed by many authors from different perspectives [4-13].

In this paper, we construct a class of coherent states for the SPT potential called \emph{epsilon} coherent states ($\varepsilon$-CS),
labeled by points $z$ of the complex plane and
depending on three numbers $(\gamma, \nu)$ and $\varepsilon>0$. Precisely, an $\varepsilon$-CS belongs to the states Hilbert space $\mathcal{H}:=L^2([0,\pi])$ of the
Hamiltonian with the SPT potential and  is defined as a superposition of eigenstates of the Hamiltonian which is encoded by $\nu$.
In this superposition, the role of coefficients
$z^n/\sqrt{n!}$ of the canonical coherent states is played by coefficients with the form  $\ e^{in\arg
z}L_n^{(\gamma-1)}(z\bar{z})/\sqrt{\sigma_{\gamma,\varepsilon}(n)}$, where $L_n^{(\alpha)}(.)$ denotes the Laguerre polynomial and $\sigma_{\gamma,\varepsilon}(n)$ are constants given below in (4.2).
These coefficients generalize, with respect to the parameter $\varepsilon$,
 those that have been used to construct a class of coherent states for the isotonic oscillator [14].

Here, the resolution of the identity of the states Hilbert space carrying these $\varepsilon$-CS is obtained at the limit $\varepsilon\to 0^{+}$ by exploiting
a result on the Poisson kernel for Gegenbauer polynomials, which is due to Muckenhoopt and Stein [15]. The method we are using is similar to what the second author used in
previous works [16-17] and it makes possible to obtain a closed form for the constructed $\varepsilon$-CS when the parameters $\gamma$ and $\nu$ are connected in a special way. Finally, we propose a suitable  definition for the associated coherent states transform.
The latter one suggests us an integral representation of Hankel type for Laguerre functions, which constitutes a result of independent interest.

The paper is organized as follows. In section 2, we recall briefly some known facts on the Hamiltonian with the SPT potential. Section 3 is devoted
to the coherent states formalism we will be using. This formalism is applied in section 4 so as to construct a class of coherent states in the
states Hilbert space of the Hamiltonian. In section 5, we give a closed form for these
states. In section 6, we propose a definition for the associated coherent states transform which leads
 to an integral representation for Laguerre functions.
 \section{The symmetric P\"{o}schl-Teller oscillator}
The problem of finding the energy spectrum and the wavefunctions of a particle confined inside the infinite square well and
submitted to a barrier of infinite potential at both frontiers has commonly been one of the most illustrative problems of quantum mechanics.
This problem is generalized to a situation in
which the confining potential is given by a family of continuously indexed potentials of the P\"{o}schl-Teller type [3]. Precisely, one considers
the motion of a particle of mass $m_{\ast}$ living in the interval $[0, L]$ and evolving under the PT potential:
\begin{equation}
V_{\eta,\delta}(x):=\frac{1}{4}\mathcal{E}_0\left(\frac{\eta(\eta-1)}{\cos^2\frac{\pi}{2L}x}+\frac{\delta(\delta-1)}{\sin^2\frac{\pi}{2L}x}\right),\ 0\leq x\leq L,
\end{equation}
where $\eta,\delta>0$, $\mathcal{E}_0:=\hbar^2\pi^2(2m_{\ast}L^2)^{-1}$ is a coupling constant and $\hbar$ being
the Planck's constant. The SPT potential corresponds to the case $\eta=\delta$. For our purpose,
we set $\eta=\delta=\nu+1$ with $\nu>-1$. Thus, the potential in (2.1) reduces to
\begin{equation}
V_{\nu}(x):=\mathcal{E}_0\frac{\nu(\nu+1)}{\sin^2\frac{\pi}{L}x}.
\end{equation}
To this potential corresponds the Hamiltonian operator or P\"{o}schl-Teller oscillator
\begin{equation}
\triangle_{\nu}:=-\frac{\hbar^2}{2m_{\ast}}\frac{d^2}{dx^2}+V_{\nu}(x)
\end{equation}
acting on the Hilbert space $\mathcal{H}:=L^2([0,L], dx)$. Its eigenvalues are quadratic with the form
\begin{equation}
E_n=\mathcal{E}_0(n+\nu+1)^2, \ \ \ n=0,1,2,\cdots ,
\end{equation}
and the normalized eigenstates obeying the Dirichlet boundary conditions are given by
\begin{equation}
\phi_{n,L}^{\nu}(x):=\Gamma(\nu+1)\frac{2^{\nu+1/2}}{\sqrt{L}}\sqrt{\frac{n!(n+\nu+1)}{\Gamma(n+2\nu+2)}}\left(\sin
\frac{\pi}{L}x\right)^{\nu+1}C_n^{\nu+1}\left(\cos\frac{\pi}{L}x\right),
\end{equation}
where $C_n^{\nu+1}(.)$ are Gegenbauer polynomials [18].
For the sake of simplicity, we choose the units $\hbar=2m_{\ast}=1$ and $L=\pi$ so that we will be dealing with  eigenvalues
$\widetilde{E}_n:=(n+\nu+1)^2,\ n=0,1,2,\cdots,$ together with the eigenstates
\begin{equation}
\phi_n^{\nu}(x):\equiv\phi_{n,\pi}^{\nu}(x)=\Gamma(\nu+1)2^{\nu+1/2}\sqrt{\frac{n!(n+\nu+1)}{\pi\Gamma(n+2\nu+2)}}\left(\sin x\right)^{\nu+1}C_n^{\nu+1}\left(\cos x\right),
\end{equation}
which constitute a complete orthonormal basis in $L^2([0,\pi], dx)$. More detailed information on the spectral theory of
 the Hamiltonian operator in (2.3) can be found in [19].\\ \\
\textbf{Remark 2.1.}
The case $\nu\to0$ corresponds to the infinite square well whose ground state energy is represented by the factor $\mathcal{E}_{0}$. In this case, the Gegenbauer polynomial in (2.6) can
be expressed in terms of the  Tchebicheff polynomial of the second kind ([18], p.97) as follows
\begin{equation}
C_n^{1}\left(\cos L^{-1}\pi x\right)=\frac{\sin\left((n+1)L^{-1}\pi x\right)}{\sin L^{-1}\pi x}.
\end{equation}
Using (2.7) one can recover the eigenstates of the Hamiltonian with the infinite square well potential as
\begin{equation}
\phi_{n,L}^{0}=\sqrt{\frac{2}{L}}\sin\frac{(n+1)\pi}{L}x
\end{equation}
for every $n=0,1,2,\cdots$.
\section{The epsilon coherent states}

In this section, we will review a generalization of canonical coherent
states by considering a kind of the identity resolution that we obtain at
the zero limit with respect to a parameter $\varepsilon >0$. These states
will be called \emph{epsilon} coherent states and denoted $\varepsilon$-CS for
brevity. Their formalism was introduced in [16-17] where new families of coherent states attached to the pseudo-harmonic
oscillator were constructed.\\

\textbf{Definition 3.1} . \textit{Let }$\mathcal{H}$\textit{\ be a (complex, separable, infinite-dimensional)
Hilbert space with an orthonormal basis }$\left\{ \psi _{n}\right\}
_{n=0}^{\infty }.$\textit{\ Let }$\mathfrak{D}$\textit{\ }$\subseteq
\mathbb{C}$\textit{\ be an open subset of }$\mathbb{C}$\textit{\ and let }$%
c_n:\mathfrak{D}\rightarrow \mathbb{C};n=0,1,2,\cdots ,$\textit{%
\ be a sequence of complex functions. Define }
\begin{equation}
\left|z,\varepsilon \right\rangle:=\left( \mathcal{N}_{\varepsilon }\left( z\right)
\right) ^{-\frac{1}{2}}\sum\limits_{n=0}^{+\infty }\frac{\overline{c%
_{n}\left( z\right) }}{\sqrt{\sigma _{\varepsilon }\left( n\right) }}\left|\psi _{n}\right\rangle
\end{equation}
\textit{where }$\mathcal{N}_{\varepsilon }\left( z\right) $\textit{\ is a
normalization factor and }$\sigma _{\varepsilon }\left( n\right) $\textit{; }%
$n=0,1,2,\cdots ,$\textit{\ a sequence of positive numbers depending on }$%
\varepsilon >0$\textit{. The vectors} $\left\{ \left|z,\varepsilon\right\rangle,z\in \mathfrak{D}\right\} $\textit{\ are said to form a set of
epsilon coherent states if :\newline
}$\left( i\right) $\textit{\ for each fixed $z\in \mathfrak{D}$ }\textit{\ and $\varepsilon >0$}$%
,$\textit{\ the state in (3.1)} \textit{\ is normalized,
that is }$\left\langle z,\varepsilon| z,\varepsilon \right\rangle_{\mathcal{H}}=1,$\textit{%
\newline
}$\left( ii\right)$ \textit{the following resolution of the identity operator on $\mathcal{H}$ }
\begin{equation}
\lim_{\varepsilon\to 0^{+}}\int\limits_{\mathfrak{D}}
\left|z,\varepsilon \right\rangle\left\langle z,\varepsilon\right|d\mu_{\varepsilon }\left( z\right) =%
\mathbf{1}_{\mathcal{H}}
\end{equation}
\textit{is satisfied with an appropriately chosen measure }$d\mu_{\varepsilon }$.\\

In the above definition, the Dirac's\textit{\ bra-ket%
} notation $\left|z,\varepsilon \right\rangle\left\langle z,\varepsilon \right|$ means the
rank-one-operator $\varphi\mapsto\left|z,\varepsilon\right\rangle\left\langle z,\varepsilon|\varphi\right\rangle_{\mathcal{H}}$, $\varphi \in
\mathcal{H}.$ Also, the
limit in $\left(3.2\right) $ is to be understood as follows. Define the
operator
\begin{equation}
\mathcal{O}_{\varepsilon }\left[ \varphi \right] \left( \cdot \right)
:=\left( \int\limits_{\mathfrak{D}}\left|z,\varepsilon\right\rangle\left\langle z,\varepsilon\right|d\mu _{\varepsilon }\left( z\right) \right)
\left[ \varphi \right] \left(
\cdot \right).
\end{equation}
Then, the above limit means $\mathcal{O}%
_{\varepsilon }\left[ \varphi \right] \left( \cdot \right) \rightarrow $ $%
\varphi \left( \cdot \right) $ as $\varepsilon \rightarrow 0^{+},$ \textit{%
almost every where }with respect to $\left( \cdot \right) .$ \\

\textbf{Remark 3.1.} The formula $\left(3.1\right) $\ can be viewed as a
generalization of the series expansion of the canonical coherent states
\begin{equation}
\left|z\right\rangle:=\left( e^{z\bar{z}}\right) ^{-\frac{1}{2}%
}\sum_{n=0}^{+\infty }\frac{z^{n}}{\sqrt{n!}}\left|\varphi_{n}\right\rangle;\quad z\in
\mathbb{C},
\end{equation}
where $\left\{\left|\varphi _{n}\right\rangle\right\}$ is an orthonormal basis
in $L^{2}\left( \mathbb{R}\right) $, which consists of eigenstates of the harmonic
oscillator, given by $\varphi_n(y)=(\sqrt{\pi}2^nn!)^{-1/2}e^{-\frac{1}{2}y^2}H_n(y)$ in terms of the Hermite polynomial [18].
\section{Epsilon CS with Laguerre coefficients for the SPT potential}

We now construct a class of $\varepsilon$-CS indexed by points $z\in\mathbb{C}$
and depending on three numbers $(\gamma,\nu)$ and $\varepsilon$ by replacing the coefficients $%
z^{n}/\sqrt{n!}$ of the canonical coherent states by Laguerre polynomials as
mentioned in the introduction.\\

\textbf{Definition 4.1.} \emph{Define a set of states labeled by points $z \in \mathbb{C}$ and depending on three numbers
$(\gamma,\nu)$ and $\varepsilon>0$ by the following superposition}
\begin{equation}
\left|z; (\gamma,\nu);\varepsilon \right\rangle:=\left(\mathcal{N}_{\gamma,\varepsilon}(z)\right)^{-\frac{1}{2}}\sum_{n=0}^{+\infty}\frac{1}{\sqrt{\sigma_{\gamma,\varepsilon}(n)}}e^{-in\arg z}L_n^{(\gamma-1)}(z\bar{z})\left|\phi_n^{\nu}\right\rangle
\end{equation}
\emph{where $\mathcal{N}_{\gamma,\varepsilon }\left( z\right) $ is a normalization factor such that} $\left\langle z;(\gamma,\nu);\varepsilon|z;(\gamma,\nu);\varepsilon\right\rangle=1$\emph{, the coefficients}
$\sigma_{\gamma, \varepsilon}(n)$\emph{  are a sequence of positive numbers given by}
\begin{equation}
\sigma _{\gamma, \varepsilon}\left(
n\right):=\frac{1}{n!}\frac{\Gamma(\gamma+n)}{\Gamma(\gamma)}\left(\frac{\gamma}{2}+n\right)e^{n\varepsilon} ,\ \ n=0,1,2,\cdots,
\end{equation}
\emph{and} $\left\{\left|\phi _{n}^{\nu}\right\rangle\right\}$ \emph{is the orthonormal basis of} $\mathcal{H}=L^{2}\left(\left[0,\pi\right], dx\right)$ \emph{as defined in} \emph{(2.6)}.\\ \\
We shall give the main properties on these states in the following.\\ \\
\textbf{Proposition 4.1.} \textit{Let  }$\gamma,\varepsilon >0 $ \emph{and} $\nu>-1$\emph{ be fixed}
\emph{parameters.} \emph{Then, for every} $z$ \emph{and} $w$ \emph{in} $\mathbb{C}$, \emph{the overlap relation}
\emph{between two} $\varepsilon$-\emph{CS is given through the scalar product}
\begin{equation}
\left\langle z;(\gamma,\nu);\varepsilon |w;(\gamma,\nu);\varepsilon \right\rangle_{\mathcal{H}}=\left(\mathcal{N}_{\gamma,\varepsilon }\left(z\right)\mathcal{N}_{\gamma,\varepsilon
}\left( w\right)\right)^{-1/2}\Gamma(\gamma)e^{\frac{1}{2}(\gamma\varepsilon+i(1-\gamma)(\arg z-\arg w))}
(zw)^{1-\gamma}
\end{equation}
\begin{equation*}
\times\int^{e^{-\varepsilon}}_{0}\frac{1}{\sqrt{t}\left(1-te^{i(\arg z-\arg w)}\right)}\exp{\left(-\frac{\left(z\bar{z}+w\overline{w}\right)te^{i(\arg
z-\arg w)}}{1-t\ e^{i(\arg z-\arg w)}}\right)}I_{\gamma-1}\left(\frac{2|zw|\sqrt{t e^{i(\arg z-\arg w)}}}{1-t e^{i(\arg z-\arg w)}}\right)dt
\end{equation*}
\emph{where} $\mathcal{N}_{\gamma,\varepsilon}(.)$ \emph{is the normalization factor in (4.1) and} $I_{\gamma-1}(.)$ \emph{is the modified Bessel function of first}
\emph{kind and of order} $\gamma-1$.\\ \\
\textbf{Proof. }Using the orthonormality relations of the basis elements $(2.6)$, the scalar product in $\mathcal{H}$ between two $%
\varepsilon$-CS can be written as
\begin{equation}
\left\langle
z;(\gamma,\nu);\varepsilon|w;(\gamma,\nu);\varepsilon\right\rangle_{\mathcal{H}}=\left(\mathcal{N}_{\gamma,\varepsilon}(z)\right)^{-1/2}\left(\mathcal{N}_{\gamma,\varepsilon}(w)\right)^{-1/2}Q_{\gamma,\varepsilon}\left(z,w\right),
\end{equation}
where
\begin{eqnarray}
Q_{\gamma,\varepsilon}(z,w)&=&\sum_{n=0}^{+\infty}\frac{n!\Gamma(\gamma)e^{-n\varepsilon}}{\Gamma(\gamma+n)\left(\frac{\gamma}{2}+n\right)}e^{in
\arg z}L_n^{(\gamma-1)}(z\bar{z})e^{-in \arg w}L_n^{(\gamma-1)}(w\overline{w})
\end{eqnarray}
\begin{equation}
=e^{\gamma\varepsilon/2}\Gamma(\gamma)\int_{0}^{e^{-\varepsilon}}t^{\frac{\gamma}{2}-1}\left[\sum_{n=0}^{+\infty}\frac{n!}{\Gamma(\gamma+n)}\left(te^{i(arg z-arg
w)}\right)^nL_n^{(\gamma-1)}(z\bar{z})L_n^{(\gamma-1)}(w\overline{w})\right]dt.
\end{equation}
Next, we make use of the Hille-Hardy formula ([20], p.135):
\begin{equation}
\sum_{n=0}^{\infty}\frac{n!
u^n}{\Gamma(n+\alpha+1)}L_n^{(\alpha)}(\xi)L_n^{(\alpha)}(\zeta)=(1-u)^{-1}\exp{\left(-\frac{(\xi+\zeta)u}{1-u}\right)(\xi\zeta
u)^{-\alpha/2}}I_{\alpha}\left(\frac{2\sqrt{\xi\zeta u}}{1-u}\right),
\end{equation}
for the parameters $\alpha=\gamma-1, \ \xi=z\bar{z}, \ \ \zeta=w\overline{w}$ and $u=te^{i(\arg z-\arg w)}$. Therefore, (4.7) reads
\[
Q_{\gamma,\varepsilon}(z,w)=\frac{\Gamma(\gamma)e^{\gamma\varepsilon/2}}{\left(|zw|^2e^{i(\arg z-\arg
w)}\right)^{\frac{\gamma-1}{2}}}\int_{0}^{e^{-\varepsilon}}t^{-\frac{1}{2}}\left(1-te^{i(\arg z-\arg w)}\right)^{-1}
\]
\begin{equation}
\times \exp{\left(-\frac{\left(|z|^2+|w|^2\right)te^{i(\arg z-\arg w)}}{1-te^{i(\arg z-\arg w)}}\right)}
I_{\gamma-1}\left(\frac{2|zw|\left(te^{i(\arg z-\arg w)}\right)^{1/2}}{1-te^{i(\arg z-\arg w)}}\right)dt.
\end{equation}
Finally, by (4.4)-(4.8) we arrive at the result (4.3). \ \ \ \ \ \ \ \ \ \ \ \ \ \ \ \ \ \ \ \ \ \ \ \ \ \ \ \ \ \ \ \ \ \ \ \ \ \ \ \ \ \ \ \ \ \
\ \ \ \ \ \ \ \ \ \ \ \ \ \ \ \ \ \ \ \ \ \ \ \ \ \ \ \ \ \ \ \ \ \ \ \ \ \ \ \ \ \ \ \ \ \ \ \ \ $\square$\\ \\
\textbf{Corollary 4.1.}
\textit{ The normalization factor in (4.1) has the expression}
\begin{equation}
\mathcal{N}_{\gamma, \varepsilon }\left(z\right)=\Gamma(\gamma)e^{\frac{\gamma\varepsilon}{2}}
(z\bar{z})^{1-\gamma}e^{z\bar{z}}\int_{0}^{e^{-\varepsilon}}\frac{1}{(1-t)\sqrt{t}}\exp{\left(-\frac{2z\bar{z}t}{1-t}\right)}I_{\gamma-1}\left(\frac{2z\bar{z}\sqrt{t}}{1-t}\right)dt.
\end{equation}
\emph{In particular, when} $\varepsilon\to0^{+}$, \emph{this factor reduces to}
\begin{equation}
\mathcal{N}_{\gamma,\
0^{+}}(z)=e^{z\bar{z}}(z\bar{z})^{1-\gamma}I_{\frac{\gamma-1}{2}}\left(\frac{z\bar{z}}{2}\right)K_{\frac{\gamma-1}{2}}\left(\frac{z\bar{z}}{2}\right)
\end{equation}
\emph{where} $I_\nu(.)$ \emph{and} $K_\nu(.)$ \emph{are modified Bessel functions of first and second kind respectively.}
\\ \\
\textbf{Proof.}
 Putting $z=w$ in (4.3) and using the condition $\left\langle z; (\gamma,\nu);\varepsilon|z;(\gamma,\nu); \varepsilon\right\rangle=1$, we obtain the normalization factor in (4.9). When $\varepsilon\to0^{+}$, this factor becomes
\begin{equation}
\mathcal{N}_{\gamma, 0^{+} }\left(z\right)=\Gamma(\gamma)(z\bar{z})^{1-\gamma}e^{z\bar{z}}\int_{0}^{1}\frac{1}{(1-t)\sqrt{t}}\exp{\left(-\frac{2z\bar{z}t}{1-t}\right)}I_{\gamma-1}\left(\frac{2z\bar{z}\sqrt{t}}{1-t}\right)dt.
\end{equation}
Changing the variable by $\sqrt{t}=\tanh\rho$ transforms (4.11) as follows
\begin{eqnarray}
\mathcal{N}_{\gamma,0^{+}}(z)=\Gamma(\gamma)(z\bar{z})^{1-\gamma}\int^{+\infty}_0\exp{\left(-z\bar{z}(\cosh\rho-1)\right)}I_{\gamma-1}(z\bar{z}\sinh\rho)d\rho.
\end{eqnarray}
Next, recalling the relation between Bessel functions ([21], p.77):
\begin{equation}
I_{\mu}(u)=e^{-i\frac{\pi}{2}\mu}J_{\mu}(iu),
\end{equation}
then (4.12) reads
\begin{eqnarray}
\mathcal{N}_{\gamma,0^{+}}(z)=\Gamma(\gamma)e^{(1-\gamma)\frac{\pi}{2}i}(z\bar{z})^{1-\gamma}e^{z\bar{z}}\int_0^{+\infty}\exp{(-z\bar{z}\cosh\rho)}J_{\gamma-1}(iz\bar{z}\sinh\rho)d\rho.
\end{eqnarray}
We now apply the formula ([22], p.363):
\begin{equation}
\int^{+\infty}_{0}\coth^{2\kappa}\left(\frac{y}{2}\right)e^{-b\cosh y}J_{2\mu}(a\sinh y)dy=\frac{\Gamma(\frac{1}{2}-\kappa+\mu)}{a\Gamma(2\mu+1)}
M_{-\kappa,\mu}(\sqrt{a^2+b^2}-b)W_{\kappa,\mu}(\sqrt{a^2+b^2}+b),
\end{equation}
$\Re b>|\Re a|$  and $\Re(\mu-\kappa)>-\frac{1}{2}$ where  $M_{-\kappa,\mu}(.)$ and $W_{\kappa, \mu}(.)$ are Whittaker functions.
 For the parameters $\kappa=0, \ a=iz\bar{z},\ b=z\bar{z}$ and $ \mu=(\gamma-1)/2$, this gives
\begin{equation}
\mathcal{N}_{\gamma,0^{+}}(z)=-i\Gamma\left(\frac{\gamma}{2}\right)e^{(1-\gamma)\frac{\pi}{2}i}(z\bar{z})^{-\gamma}e^{z\bar{z}}M_{0,\frac{\gamma-1}{2}}(-z\bar{z})W_{0,\frac{\gamma-1}{2}}(z\bar{z}).
\end{equation}
The Whittaker functions (4.16) can also be expressed in terms of Bessel functions of
first and second kind respectively through the relations ([23], p.207):
\begin{equation}
M_{0,\frac{1}{2}\lambda}(\zeta)=\frac{1}{\Gamma\left(\frac{1+\lambda}{2}\right)}\sqrt{\pi\zeta}I_{\frac{1}{2}\lambda}(\zeta), \ \ \
W_{0, \frac{1}{2}\lambda}(\xi)=\sqrt{\frac{\xi}{\pi}}K_{\frac{1}{2}\lambda}\left(\frac{\xi}{2}\right),
\end{equation}
in the case of parameters $\zeta=-z\bar{z}$, $\xi=z\bar{z}$ and $ \lambda=\gamma-1$. Finally, by using the well known property
$I_{\mu}(-\zeta)=e^{i\pi\mu}I_{\mu}(\zeta)$ we arrive at the normalization factor in (4.10). \ \ \ \ \ \ \ \ \ \ \ \ \ \ \ \ \ \ \ \ \ \ \ \ \ \ \ \ \ \ \ $\Box$\\ \\
\textbf{Remark 4.1.} As mentioned above, when $\varepsilon\to0^{+},$ the coefficients $(\sigma_{\gamma,0^{+}}(n))^{-1/2}e^{in\arg z}L_n^{(\gamma-1)}(z\bar{z})$
turn out to be those used by the authors [14] in their construction of a set of coherent states for the isotonic oscillator $H_A:=-\partial^2_x+x^2+A/x^2,\ \ A\geq0$. The normalization factor in (4.10) is the same as in ([14], p.4) where it was referred
to a proof in [24] based on the construction of the Green's function for  $H_{A}$.\\ \\
\textbf{Proposition 4.2.} \emph{Let} $\gamma>0$.\emph{ Then, the} $\varepsilon$-CS \emph{in (4.1)} \emph{satisfy the following resolution of the identity}
\begin{equation}
\lim_{\varepsilon\to0^{+}}\int_{\mathbb{C}}\left|z;(\gamma,\nu);\varepsilon\right\rangle\left\langle \varepsilon;(\gamma,\nu);z\right|d\mu_{\gamma,\varepsilon}(z)=1_{L^2([0,\pi])},
\end{equation}
\emph{where}
\begin{equation}
d\mu_{\gamma,\varepsilon}(z):=\frac{1}{\Gamma(\gamma)}\mathcal{N}_{\gamma,\varepsilon}(z)(z\bar{z})^{\gamma}e^{-z\bar{z}}d\mu(z),
\end{equation}
$\mathcal{N}_{\gamma,\varepsilon}(z)$\emph{ being the normalization factor} \emph{in (4.9) and} $d\mu(z)$ \emph{is the Lebesgue measure on} $\mathbb{C}$.\\ \\
\textbf{Proof.} Let us assume that the measure takes the form
\begin{equation}
d\mu_{\gamma,\varepsilon}(z)=\mathcal{N}_{\gamma,\varepsilon}(z)\Omega(z\bar{z})d\mu(z),
\end{equation}
where $\Omega(z\bar{z})$ is an auxiliary density function to be determined. In terms of polar coordinates $z=\rho e^{i\theta}$, Eq.(4.20) reads
\begin{equation}
d\mu_{\gamma,\varepsilon}(z)=\mathcal{N}_{\gamma,\varepsilon}(\rho^2)\Omega(\rho^2)\rho d\rho\frac{d\theta}{2\pi}.
\end{equation}
 Let $\varphi\in L^2([0,\pi], dx)$ and let us start by writing the following action:
\begin{eqnarray}
\mathcal{O}_{\gamma, \nu,\varepsilon}[\varphi]=\int_{\mathbb{C}}\left|z;(\gamma,\nu);\varepsilon\right\rangle\left\langle \varepsilon;(\gamma,\nu);z|\varphi\right\rangle d\mu_{\gamma,\varepsilon}(z)\
\end{eqnarray}
\begin{eqnarray}
=\left(\sum_{n,m=0}^{+\infty}\left(\int_0^{+\infty}\frac{L_n^{(\gamma-1)}(\rho^2)}{\sqrt{\sigma_{\gamma,\varepsilon}(n)}}\frac{L_m^{(\gamma-1)}(\rho^2)}{\sqrt{\sigma_{\gamma,\varepsilon}(m)}}\Omega(\rho^2)\rho d\rho \int^{2\pi}_{0}e^{i(m-n)\theta}\frac{d\theta}{2\pi}\right)\left|\phi_n^{\nu}\right\rangle\left\langle \phi_n^{\nu}\right|\right)[\varphi]
\end{eqnarray}
\begin{eqnarray}
=\left(\sum_{n=0}^{+\infty}\left(\frac{1}{\sigma_{\gamma,\varepsilon}(n)}\int^{+\infty}_{0}\left(L_n^{(\gamma-1)}(\rho^2)\right)^2\Omega(\rho^2)\rho d\rho\right)\left|\phi_n^{\nu}\right\rangle\left\langle \phi_n^{\nu}\right|\right)[\varphi]
\end{eqnarray}
\begin{equation}
=\left(\sum_{n=0}^{+\infty}\left(\frac{n!\Gamma(\gamma)}{\Gamma(\gamma+n)\left(2n+\gamma\right)}\int^{+\infty}_{0}\left(L_n^{(\gamma-1)}(r)\right)^2\Omega(r) dr\right)e^{-n\varepsilon}\left|\phi_n^{\nu}\right\rangle\left\langle \phi_n^{\nu}\right|\right)[\varphi].
\end{equation}
Now, we need to determine $\Omega(r)$ such that
\begin{equation}
\int_0^{+\infty}\left(L_n^{(\gamma-1)}(r)\right)^2\Omega(r)dr=\frac{(\gamma)_n}{n!}\left(2n+\gamma\right),
\end{equation}
where $(\gamma)_n:=\Gamma(\gamma+n)/\Gamma(\gamma)$ is the shifted factorial.
To do that, we make appeal to the integral  ([25], p.1133):
\begin{equation}
\int_{0}^{+\infty}y^{\sigma}e^{-y}L_k^{(\alpha)}(y)L_m^{(\beta)}(y)=(-1)^{k+m}\Gamma(\sigma+1)\sum_{j=0}^{\min(k,m)}\left(
                                                                                                         \begin{array}{c}
                                                                                                           \sigma-\alpha \\
                                                                                                           k-j \\
                                                                                                         \end{array}
                                                                                                       \right)\left(
                                                                                                                \begin{array}{c}
                                                                                                                  \sigma-\beta \\
                                                                                                                  m-j \\
                                                                                                                \end{array}
                                                                                                              \right)\left(
                                                                                                                       \begin{array}{c}
                                                                                                                         \sigma+j \\
                                                                                                                         j \\
                                                                                                                       \end{array}
                                                                                                                     \right)
\end{equation}
with conditions $\Re (\sigma)>-1$. In the case of parameters $\ y=r,\ \alpha=\beta=\sigma-1$ and $k=m=n$, this integral reduces to
\begin{equation}
\int^{+\infty}_{0}r^{\alpha+1}e^{-r}\left(L_n^{(\alpha)}(y)\right)^2dr=\frac{\Gamma(n+\alpha+1)}{n!}(2n+\alpha+1).
\end{equation}
For the parameter $\alpha=\gamma-1$, (4.28) reads
\begin{equation}
\int^{+\infty}_{0}r^{\gamma}e^{-r}\left(L_n^{(\gamma-1)}(y)\right)^2dr=\Gamma(\gamma)\frac{(\gamma)_n}{n!}(2n+\gamma).
\end{equation}
This suggests us to take
\begin{equation}
\Omega(r):=r^{\gamma}e^{-r}.
\end{equation}
Therefore, the measure in (4.20) has the form
\begin{equation}
d\mu_{\gamma,\varepsilon}(z)=\frac{1}{\Gamma(\gamma)}\mathcal{N}_{\gamma,\varepsilon}(z)(z\bar{z})^{\gamma}e^{-z\bar{z}}d\mu(z).
\end{equation}
With this measure Eq.(4.25) reduces to
\begin{equation}
\mathcal{O}_{\gamma,\nu,\varepsilon}[\varphi]\equiv\mathcal{O}_{\nu,\varepsilon}[\varphi]=\sum_{n=0}^{+\infty} e^{-n\varepsilon}\left|\phi_n^{\nu}\right\rangle\left\langle \phi_n^{\nu}\right|,
\end{equation}
for every $\varphi\in L^2([0,\pi], dx)$. Now, for $x\in[0,\pi]$, we can write
\begin{eqnarray}
\mathcal{O}_{\nu,\varepsilon}[\varphi](x)&=&\sum_{n=0}^{+\infty}e^{-n\varepsilon}\left\langle \varphi|\phi_n^{\nu}\right\rangle_{\mathcal{H}}\left\langle x|\phi_n^{\nu}\right\rangle\\
&=&\sum_{n=0}^{+\infty}e^{-n\varepsilon}\left(\int_0^{\pi}\varphi(y)\overline{\phi_n^{\nu}(y)}dy\right)\phi_n^{\nu}(x)\\
&=&\int_{0}^{\pi}\left(\sum_{n=0}^{+\infty}\left(e^{-\varepsilon}\right)^n\phi_n^{\nu}(x)\overline{\phi_n^{\nu}(y)}\right)\varphi(y)dy.
\end{eqnarray}
Replacing the $\left\{\phi_n^{\nu}\right\}$ by their expressions in (2.6), Eq.(4.35) becomes
\begin{equation}
\left(\sin x\right)^{-\nu-1}\mathcal{O}_{\nu,\varepsilon}[\varphi](x)=\int_0^{\pi}\left(\sum_{n=0}^{+\infty}e^{-n\varepsilon}\omega_nC_n^{\nu+1}(\cos x)C_n^{\nu+1}(\cos y)\right)h(y)dm_{(\nu+1)}(y),
\end{equation}
where
\begin{equation}
\omega_n:=\Gamma^2(\nu+1)2^{2\nu+1}\frac{n!(n+\nu+1)}{\pi\Gamma(n+2\nu+2)},
\end{equation}
$h(y):=\left(\sin x\right)^{-\nu-1}\varphi(y)$ and $dm_{(\nu+1)}(y)=(\sin y)^{2(\nu+1)}dy$.
In the above formal calculations (4.34)-(4.35), reversing the order of summation and integration can be justified in the more general setting of Jacobi
polynomials, see for example ([26], pp.3-4). We recognize in (4.36) the Poisson kernel for Gegenbauer polynomial ([15], p.25):
\begin{equation}
P_{\nu}(e^{-\varepsilon}; x, y)=\sum_{n=0}^{+\infty}\left(e^{-\varepsilon}\right)^n\omega_n C_n^{\nu+1}(x)C_n^{\nu+1}(y).
\end{equation}
This kernel function can also be written in a closed form by making use of the Bailey formula ([27], p.102). With these notations, Eq.(4.36) reads
\begin{equation}
(\sin x)^{-\nu-1}\mathcal{O}_{\nu,\varepsilon}[\varphi](x)=\int_0^{\pi}P_{\nu}\left(e^{-\varepsilon};x,y\right)h(y)dm_{(\nu+1)}(y).
\end{equation}
Direct calculations show that $h\in L^2([0,\pi],dm_{\nu+1}(y))$ and its norm is given by
\begin{equation}
\left\|h\right\|_{\left(L^2([0,\pi]),\ dm_{(\nu+1)}(y)\right)}=\left\|\varphi\right\|_{L^2([0,\pi])}.
\end{equation}
Therefore, by applying a result due to Muchenhoopt and Stein ([15], Theorem.2, (c), $p=2$), the right hand side of (4.39) can be denoted by $h(e^{-{\varepsilon}},x)$, and leads to the fact
\begin{equation}
\lim_{\varepsilon\to0^{+}}\left(\sin x\right)^{-\nu-1}\mathcal{O}_{\nu,\varepsilon}[\varphi](x)=\lim_{\varepsilon\to0^{+}}h(e^{-\varepsilon},x)=h(x)=\left(\sin x\right)^{-\nu-1}\varphi(x),\ \ a.e.,
\end{equation}
which says that the limit $ \mathcal{O}_{\nu,\varepsilon}[\varphi](x)\to\varphi(x), \ a.e.$ as $\varepsilon\to0^{+}$
is valid for every  $\varphi\in L^2\left([0,\pi]\right)$. In other words,
\begin{equation}
\lim_{\varepsilon\to0^{+}}\int_{0}^{\pi}\left|z;(\gamma,\nu);\varepsilon\right\rangle\left\langle z;(\gamma,\nu);\varepsilon\right|d\mu_{\gamma,\varepsilon}(z)=1_{L^2([0,\pi])}.
\end{equation}
This completes the proof. \ \ \ \ \ \ \ \ \ \ \ \ \ \ \ \ \ \ \ \ \ \ \ \ \ \ \ \ \ \ \ \ \ \ \ \ \ \ \ \ \ \ \ \ \ \ \ \ \ \ \ \ \ \ \ \ \ \ \ \ \ \ \ \ \ \ \ \ \ \ \ \ \ \ \ \ \ \ \ \ \ \ \ \ \ \ \ \ \ \ \ \ \ \ \ \ \ \ \ \ \ \ \ \ \ \ \ \ \ \ \ \ \ \ \ \ \ \ \ \ \ \ \ \ \ \ $\square$
\section{ A closed form for the $\varepsilon$-CS}
We now assume that the parameter $\gamma$ occurring Definition.4.1 is of the form $\gamma=2(\nu+1)$ where $\nu$ is the parameter encoding the SPT potential $V_{\nu}(x)$ in (2.2).\\ \\
\textbf{Proposition 5.1.} \emph{Let }$\gamma=2(\nu+1)$ \emph{with }$\nu>0$ \emph{and} $\varepsilon>0$ \emph{be fixed parameters.} \emph{Then, the wavefunctions of the} $\varepsilon$-CS \emph{in}\emph{ (4.1)} \emph{can be written in a closed form as}
\begin{equation*}
\left\langle x|z;\nu,\varepsilon \right\rangle=\left(\mathcal{N}_{2(\nu+1),\varepsilon}(z)\right)^{-1/2}\sqrt{\Gamma(2(\nu+1))}(iz\bar{z}e^{-(i\theta+\varepsilon/2)})^{-(\nu+1/2)}(\sin x)^{1/2}
\end{equation*}
\begin{equation*}
\times \left(1-2e^{-(i\theta+\varepsilon/2)}\cos x+e^{-(2i\theta+\varepsilon)}\right)^{-1/2}\exp\left(\frac{-z\bar{z}e^{-(i\theta+\varepsilon/2)}\left(\cos x-e^{-(i\theta+\varepsilon/2)}\right)}{1-2e^{-(i\theta+\varepsilon/2)}\cos x+e^{-(2i\theta+\varepsilon)}}\right)
\end{equation*}
\begin{equation}
\times I_{\nu+\frac{1}{2}}\left(\frac{iz\bar{z}e^{-(i\theta+\varepsilon/2)}\sin x}{1-2e^{-(i\theta+\varepsilon/2)}\cos x+e^{-(2i\theta+\varepsilon)}}\right),
\end{equation}
\emph{for every} $x\in[0,\pi].$\\ \\
\textbf{Proof.} We start by the expression of the wavefunction
\begin{equation}
\left\langle x|z;\nu;\varepsilon \right\rangle:=\left( \mathcal{N}_{2(\nu+1),\varepsilon }\left(z\right)
\right) ^{-\frac{1}{2}}\sum\limits_{n=0}^{+\infty }\frac{e^{-in\arg z}L_n^{(2\nu+1)}(z\bar{z})}{\sqrt{\sigma_{2(\nu+1),\varepsilon}\left( n\right) }}\left\langle x|\phi^{\nu}_{n}\right\rangle.
\end{equation}
To get a closed form of the series
\begin{equation}
\mathcal{S}(x)=\sum_{n=0}^{+\infty}\frac{\sqrt{n!\Gamma(2(\nu+1))}e^{-in(\theta+\varepsilon/2) }}{\sqrt{\Gamma(n+2(\nu+1))\left(n+\nu+1\right)}}L_n^{(2\nu+1)}(z\bar{z})\left\langle x|\phi^{\nu}_{n}\right\rangle
\end{equation}
we replace $\left\langle x|\phi^{\nu}_{n}\right\rangle$ by its expression in (2.6) and we use the Legendre's duplication formula ([28], p.5):
\begin{equation}
\Gamma(2\xi)=\frac{1}{\sqrt{\pi}}2^{2\xi-1}\Gamma(\xi)\Gamma\left(\xi+\frac{1}{2}\right)
\end{equation}
for $\xi=\nu+1$. Then, (5.3) reads
\begin{equation}
\mathcal{S}(x)=\frac{\sqrt{\Gamma(2(\nu+1))}}{2^{\nu+1/2}\Gamma\left(\nu+\frac{3}{2}\right)}(\sin x)^{\nu+1}\sum_{n=0}^{+\infty}\frac{n!e^{-n(i\arg z+\varepsilon/2)}}{(2(\nu+1))_n}L_n^{(2\nu+1)}(z\bar{z})C_n^{\nu+1}(\cos x).
\end{equation}
We introduce the variable $\tau:=e^{-(i\theta +\varepsilon/2)}$ and we rewrite (5.5) as follows
\begin{equation}
\mathcal{S}(x)=2^{-(\nu+1/2)}\frac{\sqrt{\Gamma(2(\nu+1))}}{\Gamma\left(\nu+\frac{3}{2}\right)}(\sin x)^{\nu+1}\mathcal{G}(x),
\end{equation}
where
\begin{equation}
\mathcal{G}(x)=\sum_{n=0}^{+\infty}\frac{n!\tau^n}{(2(\nu+1))_n}L_n^{(2\nu+1)}(z\bar{z})C_n^{\nu+1}(\cos x).
\end{equation}
We now make use of generating relation ([29], p.56):
\begin{equation}
\sum_{n=0}^{+\infty}\frac{n!v^n}{(2\eta)_n}L_n^{(2\eta-1)}(w)C_n^{\eta}(u)=\frac{1}{\left(1-2uv+v^2\right)^{\eta}}\exp\left(\frac{-wv(u-v)}{1-2uv+v^2}\right)\ _{0}F_1\left(
                                                                \begin{array}{cc}
                                                                  -; & \frac{w^2v^2(u^2-1)}{4\left(1-2uv+v^2\right)^2} \\
                                                                  \eta+\frac{1}{2}; &  \\
                                                                \end{array}
                                                              \right)
\end{equation}
for the parameters $\eta=\nu+1,\ v=\tau,\ w=z\bar{z}$ and $u=\cos x$. This gives
\begin{equation*}
\mathcal{G}(x)=\left(1-2e^{-(i\theta+\varepsilon/2)}\cos x+e^{-(2i\theta+\varepsilon)}\right)^{-(\nu+1)}\exp\left(\frac{-z\bar{z}e^{-(i\theta+\varepsilon/2)}\left(\cos x-e^{-(i\theta+\varepsilon/2)}\right)}{1-2e^{-(i\theta+\varepsilon/2)}\cos x+e^{-(2i\theta+\varepsilon)}}\right)
\end{equation*}
\begin{equation}
\times\ _0F_1\left(
                                                                               \begin{array}{cc}
                                                                                 -; & \frac{1}{4}\frac{(iz\bar{z})^2e^{-(2i\theta+\varepsilon)}\sin^2 x}{\left(1-2e^{-(i\theta+\varepsilon/2)}\cos x+e^{-(2i\theta+\varepsilon)}\right)^2} \\
                                                                                 \nu+3/2 &  \\
                                                                               \end{array}
                                                                             \right).
\end{equation}
Next, we apply the formula ([30], p.56):
\begin{equation}
\left(\frac{1}{2}y\right)^{\mu}\ _0F_1(\mu+1;\frac{1}{4}y^2)=\Gamma(\mu+1)I_{\mu}(y).
\end{equation}
for $\mu=\nu+1/2$ and $y=\left(1-2e^{-(i\theta+\varepsilon/2)}\cos x+e^{-(2i\theta+\varepsilon)}\right)^{-1}iz\bar{z}e^{-(i\theta+\varepsilon/2)}\sin x$. Therefore, Eq.(5.9) becomes
\small{\begin{equation*}
\mathcal{G}(x)=2^{\nu+\frac{1}{2}}\Gamma\left(\nu+\frac{3}{2}\right)\left(1-2e^{-(i\theta+\varepsilon/2)}\cos x+e^{-(2i\theta+\varepsilon)}\right)^{-1/2}\exp\left(\frac{-z\bar{z}e^{-(i\theta+\varepsilon/2)}\left(\cos x-e^{-(i\theta+\varepsilon/2)}\right)}{1-2e^{-(i\theta+\varepsilon/2)}\cos x+e^{-(2i\theta+\varepsilon)}}\right)
\end{equation*}}
\begin{equation}
\times \left(iz\bar{z}e^{-(i\theta+\varepsilon/2)}\right)^{-(\nu+1/2)}(\sin x)^{-(\nu+1/2)} I_{\nu+\frac{1}{2}}\left(\frac{iz\bar{z}e^{-(i\theta+\varepsilon/2)}\sin x}{1-2e^{-(i\theta+\varepsilon/2)}\cos x+e^{-(2i\theta+\varepsilon)}}\right).
\end{equation}
We arrive at
\begin{equation*}
\mathcal{S}(x)=\sqrt{\Gamma(2(\nu+1))}(iz\bar{z}e^{-(i\theta+\varepsilon/2)})^{-(\nu+1/2)}\sqrt{\sin x}\left(1-2e^{-(i\theta+\varepsilon/2)}\cos x+e^{-(2i\theta+\varepsilon)}\right)^{-1/2}
\end{equation*}
\begin{equation}
\times \exp\left(\frac{-z\bar{z}e^{-(i\theta+\varepsilon/2)}\left(\cos x-e^{-(i\theta+\varepsilon/2)}\right)}{1-2e^{-(i\theta+\varepsilon/2)}\cos x+e^{-(2i\theta+\varepsilon)}}\right)I_{\nu+\frac{1}{2}}\left(\frac{iz\bar{z}e^{-(i\theta+\varepsilon/2)}\sin x}{1-2e^{-(i\theta+\varepsilon/2)}\cos x+e^{-(2i\theta+\varepsilon)}}\right),
\end{equation}
which gives the expression (5.1).$\ \ \ \ \ \ \ \ \ \ \ \ \ \ \ \  \ \ \ \ \ \ \ \ \ \ \ \ \ \ \   \ \ \ \ \ \ \ \ \ \ \ \ \ \ \   \ \ \ \ \ \ \ \ \ \ \ \ \ \ \   \ \ \ \ \ \ \ \ \ \ \ \ \ \ \   \ \ \ \ \ \ \ \ \ \ \ \ \ \ \   \ \ \ \ \ \ \ \ \ \ \ \ \ \Box$\\ \\
\textbf{Corollary 5.1.} \emph{The wavefunctions of the} $\varepsilon$-\emph{CS for the infinite square well potential are of the form}
\begin{equation*}
\left\langle x|z;0^{+};\varepsilon \right\rangle=\left(\mathcal{N}_{\varepsilon}(z)\right)^{-1/2}\sqrt{\frac{2}{\pi}}\left(z\bar{z}e^{-(i\theta+\varepsilon/2)}\right)^{-1}\exp\left(\frac{-z\bar{z}e^{-(i\theta+\varepsilon/2)}\left(\cos x-e^{-(i\theta+\varepsilon/2)}\right)}{1-2e^{-(i\theta+\varepsilon/2)}\cos x+e^{-(2i\theta+\varepsilon)}}\right)
\end{equation*}
\begin{equation}
\times \sin\left(\frac{z\bar{z}e^{-(i\theta+\varepsilon/2)}\sin x}{1-2e^{-(i\theta+\varepsilon/2)}\cos x+e^{-(2i\theta+\varepsilon)}}\right)
\end{equation}
\emph{for every }$x\in[0,\pi]$, \emph{where} $\theta=\arg z.$\\ \\
\textbf{Proof.}
As mentioned in Remark 2.1., when $\nu\to0^{+}$ the SPT potential becomes the infinite square well potential with eigenfunctions $\phi_{n,\pi}^{0}(x)$ that one obtains
by putting $L=\pi$ in (2.8). So that the result is deduced by setting $\nu=0$ in the expression (5.1). This yields
\begin{equation*}
\left\langle x|z;0^{+};\varepsilon \right\rangle=\left(\mathcal{N}_{\varepsilon}(z)\right)^{-1/2}\sqrt{\Gamma{(2)}}(iz\bar{z}e^{-(i\theta+\varepsilon/2)})^{-1/2}(\sin x)^{1/2}\left(1-2e^{-(i\theta+\varepsilon/2)}\cos x+e^{-(2i\theta+\varepsilon)}\right)^{-1/2}
\end{equation*}
\begin{equation*}
\times \exp\left(\frac{-z\bar{z}e^{-(i\theta+\varepsilon/2)}\left(\cos x-e^{-(i\theta+\varepsilon/2)}\right)}{1-2e^{-(i\theta+\varepsilon/2)}\cos x+e^{-(2i\theta+\varepsilon)}}\right)I_{\frac{1}{2}}\left(\frac{iz\bar{z}e^{-(i\theta+\varepsilon/2)}\sin x}{1-2e^{-(i\theta+\varepsilon/2)}\cos x+e^{-(2i\theta+\varepsilon)}}\right).
\end{equation*}
By applying the well known identity $I_{p}(\xi)=i^{-p}J_{1/2}(\xi)$ for $p=1/2$ and by using the relations $J_{1/2}(\xi)=\sqrt{2}(\pi \xi)^{-1/2}\sin \xi$ see ([20], p.88), we deduce
\begin{equation}
I_{\frac{1}{2}}(i\xi)=\sqrt{\frac{2i}{\pi\xi}}\sin\xi,
\end{equation}
and obtain the result.\ \ \ \ \ \ \ \ \ \ \ \ \ \ \ \ \ \ \ \ \ \ \ \ \ \ \ \ \ \ \ \ \ \ \ \ \ \ \ \ \ \ \ \ \ \ \ \ \ \ \ \ \ \ \ \ \ \ \ \ \ \ \ \ \ \ \ \ \ \
\ \ \ \ \ \ \ \ \ \ \ \ \ \ \ \ \ \ \ \ \ \ \ \ \ \ \ \ \ \ \ \ \ \ \ \ \ \ \ \ \ \ \ \ \ \ \ \ \ \ \ \ \ \ \ \ \ \ \ \ \ \ \ \ \ \ \ \ \ \ $\Box$
\section{An integral representation for Laguerre functions}
Naturally, once we have obtained a closed form for the $\varepsilon$-CS we can look for the associated coherent state transform.
In view of (4.1), this transform should map the space $L^2([0,\pi], dx)$ spanned by eigenstates $\left\{\phi_n^{\nu}\right\}$ in (2.6) onto a subspace of the space of complex-valued square integrable functions with respect to the measure
\begin{equation}
d\lambda_{\nu}(z)=\frac{1}{\Gamma(2(\nu+1))}(z\bar{z})^{2(\nu+1)}e^{-z\bar{z}}d\mu(z),
 \end{equation}
where $d\mu(z)$ is the Lebesgue measure on $\mathbb{C}$.
Recalling that the resolution of the identity, which usually ensures the isometry property of such map, was obtained at the limit $\varepsilon\to0^{+}$ in (4.18),
 then a convenient definition for such a transform could be as follows.\\

\textbf{Definition 5.1.} \emph{Let} $\nu>0$ \emph{be  a fixed parameter}. \emph{The coherent state transform associated with the $\varepsilon$-CS in (4.1) is} \emph{the map}
$\mathcal{W}_{\nu}:L^2([0,\pi], dx)\rightarrow L^2\left(\mathbb{C},d\lambda_{\nu}(z)\right)$
\emph{defined by}
\begin{equation}
\mathcal{W}_{\nu}[\varphi](z)=\lim_{\varepsilon\to0^{+}}\sqrt{\mathcal{N}_{2(\nu+1),\varepsilon}(z)}\left\langle \varphi|z;\nu;\varepsilon\right\rangle_{\mathcal{H}},
\end{equation}
\emph{for every} $z\in\mathbb{C}$.\\ \\
In particular, according to (6.2), the image  of the basis vector $\left\{\phi^{\nu}_{n}\right\}$ under the transform $\mathcal{W}_{\nu}$ should exactly
be the coefficient: $(\sigma_{2(\nu+1),\varepsilon}(n))^{-1/2}e^{-in\arg z}L_n^{(2\nu+1)}(z\bar{z})$. More precisely,
\begin{equation*}
\lim_{\varepsilon\to0^{+}}\left(\mathcal{N}_{2(\nu+1),\varepsilon}(z)\right)^{1/2}\int_0^{\pi}\phi_n^{\nu}(x)\overline{\left\langle x|z;\nu;\varepsilon\right\rangle}dx=\frac{\sqrt{n!\Gamma(2(\nu+1))}e^{-in\arg z}}{\sqrt{\Gamma(n+2\nu+2)(n+\nu+1)}}L_n^{(2\nu+1)}(z\bar{z}).
\end{equation*}
This connection, suggests us the following result on the Laguerre function
\begin{equation}
\mathcal{L}_n^{(2\nu+1)}(x):=x^{\nu+1/2}e^{-\frac{1}{2}x}L_n^{(2\nu+1)}(x),\ \ x\geq0,
\end{equation}
which can be checked by straightforward calculations.\\ \\
\textbf{Lemma 6.1.} \emph{The following integral representation for the Laguerre function}
\begin{equation}
\mathcal{L}^{(2\nu+1)}_n(x)=\frac{2^{2\nu+2}}{\sqrt{\pi}}\Gamma(\nu+1)(n+\nu+1)\int_{0}^{+\infty}J_{\nu+\frac{1}{2}}\left(\frac{x}{2}s\right)C_n^{\nu+1}\left(\frac{s^2-1}{s^2+1}\right)\frac{s^{\nu+3/2}}{(s^2+1)^{\nu+2}}ds
\end{equation}
\emph{holds true.}\\ \\
\textbf{Proof.} We multiply the Laguerre function by $t^n$  and we use the generating function for the Laguerre polynomials ([20], p.84):
\begin{equation}
(1-t)^{-\alpha-1}\exp\left(\frac{-xt}{1-t}\right)=\sum_{n=0}^{\infty}t^nL^{(\alpha)}_n(x), \ |t|<1, \ \alpha>-1,
\end{equation}
to obtain the sum
\begin{equation}
\sum_{n=0}^{+\infty}t^{n}\mathcal{L}_n^{(2\nu+1)}(x)=(1-t)^{-2\nu-2}x^{\nu+1/2}\exp\left(-\frac{x}{2}\left(\frac{1+t}{1-t}\right)\right).
\end{equation}
On the other hand, we consider the series
\begin{equation}
S_t=\frac{2^{2\nu+2}}{\sqrt{\pi}}\Gamma(\nu+1)(n+\nu+1)\sum_{n=0}^{+\infty}t^{n}\int_{0}^{+\infty}J_{\nu+\frac{1}{2}}\left(\frac{x}{2}s\right)C_n^{\nu+1}\left(\frac{s^2-1}{s^2+1}\right)\frac{s^{\nu+3/2}}{(s^2+1)^{\nu+2}}ds
\end{equation}
which is absolutely convergent. Its sum can be obtained as follows. Reversing the order of summation and integration, gives
\begin{equation}
S_{t}=\frac{2^{2\nu+2}}{\sqrt{\pi}}\Gamma(\nu+1)\int^{+\infty}_{0}\left[\sum_{n=0}^{+\infty}(n+\nu+1)t^nC_n^{\nu+1}\left(\frac{s^2-1}{s^2+1}\right)\right]J_{\nu+1/2}\left(\frac{x}{2}s\right)\frac{s^{\nu+3/2}}{(s^2+1)^{\nu+2}}ds.
\end{equation}
By combining the two generating functions of Gengenbauer polynomials ([20], p.83) and ([31], p.449) respectively:
\begin{equation}
\sum_{n=0}^{+\infty}t^{n}C_n^{\lambda}(y)=\left(1-2yt+t^2\right)^{-\lambda}
\end{equation}
and
\begin{equation}
\sum_{n=0}^{+\infty}(n+2\lambda)t^{n}C_n^{\lambda}(y)=\frac{2\lambda(1-ty)}{\left(1-2yt+t^2\right)^{\lambda+1}}
\end{equation}
for $\lambda=\nu+1$, we obtain the identity
\begin{equation}
\sum_{n=0}^{+\infty}(n+\nu+1)t^nC_n^{\nu+1}\left(\frac{s^2-1}{s^2+1}\right)=\frac{(\nu+1)(1-t^2)(s^2+1)^{\nu+2}}{\left((1-t)^2s^2+(1+t)^2\right)^{\nu+2}}.
\end{equation}
Substituting the right hand side of (6.11) into (6.8) gives
\begin{equation}
S_t=\frac{2^{2\nu+2}}{\sqrt{\pi}}\Gamma(\nu+2)\frac{(1-t^2)}{(1-t)^{2(\nu+2)}}\int_0^{+\infty}J_{\nu+1/2}\left(\frac{x}{2}s\right)\frac{s^{\nu+3/2}}{(s^2+a_t^2)^{\nu+2}}ds
\end{equation}
where $a_t:=(1+t)(1-t)^{-1}$. The integral in (6.12) can be evaluated with the aid of the formula ([21], p.434):
\begin{equation}
\int_0^{+\infty}J_{\beta}(yu)\frac{y^{\beta+1}}{(y^2+a^2)^{\eta+1}}dy=\frac{u^{\eta}a^{\beta-\eta}}{2^{\eta}\Gamma(\eta+1)}K_{\beta-\eta}(ua),
\end{equation}
with $-1<\Re(\beta)< 2\Re(\eta)+3/2$. Indeed, for $\beta=\nu+1/2$, $\eta=\nu+1,\ u=x/2$ and $a=a_t$, in addition identity $K_{-1/2}(\zeta)=\sqrt{\pi}(2\zeta)^{-1/2}e^{-\zeta}$ see ([21], pp.79-80) we get, after some simplifications, that
\begin{equation}
S_t=(1-t)^{-2\nu-2}x^{\nu+1/2}\exp\left(-\frac{x}{2}\left(\frac{1+t}{1-t}\right)\right).
\end{equation}
Thus, for all $|t|<1$, we have established the equality
\begin{equation}
\sum_{n=0}^{+\infty}t^{n}\mathcal{L}_n^{(2\nu+1)}(x)=\sum_{n=0}^{+\infty}t^{n}\frac{\Gamma(\nu+1)(n+\nu+1)}{2^{-2\nu-2}\sqrt{\pi}}\int_{0}^{+\infty}J_{\nu+\frac{1}{2}}\left(\frac{x}{2}s\right)C_n^{\nu+1}\left(\frac{s^2-1}{s^2+1}\right)\frac{s^{\nu+3/2}}{(s^2+1)^{\nu+2}}ds,
\end{equation}
and then, comparing coefficients of identical power of $t$, we obtain the result (6.4). \ \ \ \ \ \ \ \ \ \ \ \ \ \ \ \ \ \ \ \ \ \ \ \ $\Box$\\ \\
\textbf{Acknowledgements.}
\small{At the time of the conclusion of this work on December 2014 Zouha\"{\i}r Mouayn was staying at \emph{Institut des Hautes Etudes Scientifiques}, Paris. He is grateful to IHES for hospitality and for offering him encouraging conditions to achieve his research.}
\section*{References}
\footnotesize{$[1]$ P. Harrison, Quantum Wells, Wires and Dots. Theoretical and Computational Physics, New-York: Wiley 2005\\
$[2]$ R.K. Willardson, Self-Assembled InGaAs-GaAs Quantum Dots, New York: Academic Press 1999\\
$[3]$ G. P\"{o}schl and E. Teller, \emph{Z. Physik} \textbf{83} 143 (1933)\\
$[4]$ H. Bergeron, J.-P. Gazeau, P. Siegel and A. Youssef, \emph{Eur. Phys. Lett.}\textbf{92} 60003-1-5 (2010)\\
$[5]$ H. Bergeron, P. Siegl and A. Youssef, \emph{J. Phys. A: Math. Theor.} \textbf{45} 244028 (2012)\\
$[6]$ M.G.A. Crawford and E.R. Vrscay, \emph{ Phys. Rev. A} \textbf{57} 106 (1998)\\
$[7]$ S. Cruz y Cruz, S. Kuru and J. Negro, \emph{Phys. Lett. A. }\textbf{372} 1391–1405 (2008)\\
$[8]$ A.H. El Kinani and M. Daoud, \emph{ Phys. Lett. A.} \textbf{283} 291-299 (2001)\\
$[9]$ D.J. Fern\`{a}ndez, V. Hussin and O. Rosas-Ortiz, \emph{J. Phys. A: Math. Theor.} \textbf{40} 6491-6511 (2007)\\
$[10]$ D. B. Hayrapetyan, E. M. Kazaryan and H. Kh. Tevosyan, \emph{J. Contemp. Phys.} (Armen. Acad. Sci) \textbf{49} (3) 119-122 (2014)\\
$[11]$ M. N. Hounkounnou, S. Arjika, E. Baloitcha, \emph{J. Math. Phys.} \textbf{55} 123502 (2014)\\
$[12]$ U. Roy, J.Banerji and P.K. Panigrahi, \emph{J. Phys. A: Math. Gen.} \textbf{38} 9115 (2005) \\
$[13]$ R. Sarath and P.C. Vinodkumar, \emph{Pramana-J. Phys.} 10.1007/s12043-014-0890-7 (2014)\\
$[14]$ K. Thirulogasanthar and N. Saad, \emph{J. Phys.A-Math Gen} \textbf{37} (16)  4567-4577 (2004)\\
$[15]$ B. Muchenhoupt and E. M. Stein, \emph{Trans. Amer. Math. Soc.} \textbf{118} 17-92 (1965)\\
$[16]$ Z. Mouayn, \emph{J. Phys. A: Math. Theor.} \textbf{43} 295201 (2010)\\
$[17]$ Z. Mouayn, \emph{J. Math. Phys} \textbf{53} 012103 (2012)\\
$[18]$ Mourad E.H.Ismail, Classical and Quantum Orthogonal Polynomials in one Variable, Encyclopedia of Mathematics and its
Applications, Cambridge University Press 2005\\
$[19]$ F. Gestesy and W. Kirsch, \emph{J. Rein. Ang. Math.} \textbf{362} 28-50 (1985)\\
$[20]$ H. Srivastava and L. Manocha, A Treatise on Generating Functions, Ellis Horwood Ltd, London 1984\\
$[21]$ G.N. Watson, A treatise on the theory of Bessel function, Univ. Press, Cambridge 1944 \\
$[22]$ W. Magnus, F. Oberhettinger and F.G. Tricomi, Tables of Integral Transforms, Vol \textbf{2}, McGraw-Hill Book Company, New-York 1954\\
$[23]$ H. Buchholz,The Confluent Hypergeometric Function, Vol \textbf{15}, Springer-Verlag Berlin Heildelberg  1969\\
$[24]$ A. B. von Keviczky, N. Saad and R. L. Hall, \emph{J. Math. Phys.} \textbf{46} 073512 (2005)\\
$[25]$ H. M. Srivastava, H. A. Mavromatis and R.S. Alassar, \emph{Appl. Math. Lett.} \textbf{16} 1131-1136 (2003)\\
$[26]$ C. P. Calder\'{o}n and W. O. Urbina, \emph{ Illinois. J. Math.} \textbf{57} (2) 343-371 (2013)\\
$[27]$ W.N. Bailey, Generalized Hypergeometric Series, Cambridge Tracts in Mathematics and Mathematical Physics, New York 1964\\
$[28]$ W. Magnus, F. Oberhettinger and F.G. Tricomi, Higher Transcendental Function, Vol \textbf{1}, McGraw-Hill Book Company, New-York 1953\\
$[29]$ E. B. McBride, Obtaining Generating Functions, Springer-Verlag Berlin Heidelberg, New York 1971\\
$[30]$ W. Magnus, F. Oberhettinger and F.G. Tricomi, Higher Transcendental Functions, Vol\textbf{ 2}, McGraw-Hill Book Company, Inc 1953\\
$[31]$ F. W. J. Olver, D. W. Lozier, R. F. Boisvert and C. W. Clark, NIST Handbook of Mathematical function, Cambridge University Press 2010\\ \\
${}^{\flat}$\emph{E-mail address}: kayupepatrick@gmail.com\\
${}^{\natural}$\emph{E-mail address}: mouayn@gmail.com}
\end{document}